# Precessing ball solitons as dissipative structures during a phase transition in a ferromagnet

## V.V.Nietz

*Joint Institute for Nuclear Research,*
*Dubna, Moscow region 141980, Russia*

**Abstract**

Precessing ball solitons (PBS) in a ferromagnet during the first order phase transition induced by a magnetic field directed along the axis of anisotropy, while the action of the periodic field perpendicular to the main magnetic field, have been analyzed.

Under this condition, the characteristics of arising equilibrium PBS's are uniquely determined by the frequency of the periodic field, but the solitons with other frequencies are impossible. It is shown that the equilibrium PBS's are essentially the "dissipative structures" that can arise in a metastable state.

## Introduction

A phase transition in a ferromagnet under the action of a magnetic field along the easy axis has been considered in [1]. In such cases, in the metastable state of the crystal, precessing magnetic solitons (PBS) with the symmetry of the ellipsoid can spontaneously arise. The precession frequency and, correspondingly, the amplitude, size and energy of PBS are characterized by a continuous spectrum. Under certain conditions, PBS can grow and transform into the macroscopic domain of a new phase state. This transformation is related to the energy dissipation and accompanied by a continuous change in the precession frequency. However, because of spontaneous origin of PBS with different precession frequencies and the consequent uncertainty of the other characteristics of solitons, the experimental study of these objects is complex enough.

In this paper we consider the conditions when, in addition to the main magnetic field that provides phase transition, a periodic magnetic field perpendicular to the easy axis operates. In such a case, only solitons with the precession frequency equal to the frequency of the magnetic field could arise, and, consequently, the characteristics of PBS are clearly defined, making it easier to carry out experimental research.

## Equation for PBS

In the given article, to analyze magnetic solitons in a ferromagnet at the first-order transition in the presence of periodic magnetic field, as in [1], we use the equation

$$\frac{\partial \mathbf{m}}{\partial t} = -\gamma \mathbf{m} \times \tilde{\mathbf{H}} + \kappa \left( \mathbf{m} \times \frac{\partial \mathbf{m}}{\partial t} \right) \quad (1)$$

and corresponding expression for the thermodynamic potential:

$$W = \frac{K_1}{2}|m_\perp|^2 - m_z H_z - (\mathbf{m}\mathbf{H}_\perp) + \frac{\alpha}{2}\left[\left(\frac{\partial \mathbf{m}}{\partial X}\right)^2 + \left(\frac{\partial \mathbf{m}}{\partial Y}\right)^2\right] + \frac{\alpha_z}{2}\left(\frac{\partial \mathbf{m}}{\partial X}\right)^2 -$$
$$- m_z(H_z + 4\pi M_0) + \frac{2\pi}{3}(1+m_z)^2 M_0 \quad (2)$$

Here $\mathbf{H}_\perp$ is the periodic field directed perpendicular to the Z-axis, $\mathbf{m}$ is a non-dimensional vector of ferromagnetism equal in the absolute value to $1$; $m_\perp = m_x + im_y$; $K_1 > 0$, $\kappa > 0$,

$$\gamma = \frac{2\mu_B}{\hbar}; \quad \tilde{\mathbf{H}} = -\frac{\partial W}{\partial \mathbf{m}} + \frac{\partial}{\partial X}\frac{\partial W}{\partial\left(\frac{\partial \mathbf{m}}{\partial X}\right)} + \frac{\partial}{\partial Y}\frac{\partial W}{\partial\left(\frac{\partial \mathbf{m}}{\partial Y}\right)} + \frac{\partial}{\partial Z}\frac{\partial W}{\partial\left(\frac{\partial \mathbf{m}}{\partial Z}\right)}.$$

We consider the PBS in a flat plate perpendicular to the Z-axis. Next, we use the following dimensionless values: $\tau = 2\mu_B K_1 \hbar^{-1} t$, $x = K_1^{0.5}\alpha^{-0.5} X$, $y = K_1^{0.5}\alpha^{-0.5} Y$, $z = K_1^{0.5}\alpha_z^{-0.5} Z$; $h = H_z/K_1$.

If the added periodic field is

$$\mathbf{H}_\perp = K_1 h_\perp e^{-i\omega\tau} \quad (3)$$

and to present the expression for magnetic component in the view

$$m_\perp(r,\tau) = p(r,\tau) e^{-i(\omega\tau + \beta(r,\tau))}, \quad (4)$$

we obtain the following equations for $m_z(r,\tau)$:

$$\frac{\partial^2 m_z}{\partial r^2} + \frac{2}{r}\frac{\partial m_z}{\partial r} + \frac{m_z}{1-m_z^2}\left(\frac{\partial m_z}{\partial r}\right)^2 = -(1-m_z^2)\left[(1-D)m_z + \left(h - D + \omega + \frac{\partial \beta}{\partial \tau}\right)\right] +$$
$$+ \kappa \frac{\partial m_z}{\partial \tau} + h_\perp m_z \sqrt{1-m_z^2}\cos\beta - m_z\sqrt{1-m_z^2}\left(\frac{\partial \beta}{\partial r}\right)^2 \quad (5)$$

$$\frac{\partial m_z}{\partial \tau} = -\sqrt{1-m_z^2}\left[\kappa\sqrt{1-m_z^2}\left(\omega + \frac{\partial \beta}{\partial \tau}\right) + h_\perp \sin\beta\right] + \frac{1}{2i}\nabla(m\nabla m^* - m^*\nabla m). \quad (6)$$

In (5) and further, we note: $h = h_z + 4\pi M_0/K_1$, $D = 4\pi M_0/3K_1$.

The phase of precession of magnetic moments for localized excitation differs from the phase of periodic field and depends on a radius, i.e. $\beta = \beta(r,\tau)$.

For the density of PBS energy (together with the energy of interaction with the external field), relative to the initial state, we have the expression (see [1]):

$$e(r,\tau) = \frac{(1-D)(1-m_z^2)}{2} - (h-D)m_z + \frac{1}{2(1-m_z^2)}\left(\frac{\partial m_z}{\partial r}\right)^2 - h_\perp\sqrt{1-m_z^2}\cos\beta - h + D. \quad (7)$$

The change of this energy is connected with dissipation and the action of the external periodic field and equals

$$\left(\frac{de(r,\tau)}{d\tau}\right)_{change} = -\kappa\left[\frac{1}{1-m_z^2}\left(\frac{\partial m_z}{\partial \tau}\right)^2 + (1-m_z^2)\left(\omega + \frac{\partial \beta}{\partial \tau}\right)^2\right] - h_\perp\sqrt{1-m_z^2}\,\omega\sin\beta. \quad (8)$$

The equations (5)–(8) constitute a complete description of PBS, including their time transformation. (Here, we are not taking into account the spatial movement of PBS.) However, in the given paper we consider the equilibrium state of PBS inside the ferromagnet, i.e. when the decrease of energy caused by dissipation is compensated by energy flow from the external periodic field, i.e.

$$\left(\frac{de(r,\tau)}{d\tau}\right)_{change} = -\omega\sqrt{1-m_z^2}\left(\kappa\omega\sqrt{1-m_z^2} + h_\perp\sin\beta\right) = 0. \quad (9)$$

In such case

$$\frac{\partial p}{\partial \tau} = \frac{\partial m_z}{\partial \tau} = \frac{\partial \beta}{\partial \tau} = 0 \ . \quad (10)$$

Therefore, for the equilibrium state of PBS:

$$\beta(r) = -\arcsin\left(a\sqrt{1-m_z(r)^2}\right), \quad (11)$$

where $a = \kappa\omega/h_\perp$.

Corresponding equation for PBS is the following:

$$\frac{d^2 m_z}{dr^2} + \frac{2}{r}\frac{dm_z}{dr} + \frac{m_z}{1-m_z^2}\left(1 + \frac{a^2 m_z^2(1-m_z^2)}{1-a^2(1-m_z^2)}\right)\left(\frac{dm_z}{dr}\right)^2 =$$
$$= -(1-m_z^2)[(1-D)m_z + (h-D+\omega)] + h_\perp m_z\sqrt{1-m_z^2}\sqrt{1-a^2(1-m_z^2)} \quad (12)$$

Considering equilibrium PBS's are "dissipative structures" [2, 3] that can origin spontaneously in a metastable state during the first-order phase transition to the stable equilibrium state, i.e. to $m_z = +1$. For such equilibrium PBS, the entropy increase connected with dissipation is compensated by the decrease of the entropy that due the external periodic field. It can be expressed as follows:

$$\frac{dS}{d\tau} = \frac{dS_{diss}}{d\tau} + \frac{dS_{h_\perp}}{d\tau} = 0, \quad (13)$$

where $\dfrac{dS_{diss}}{d\tau} = -\dfrac{dS_{h_\perp}}{d\tau} = \dfrac{\kappa \omega^2}{T}\int_0^\infty (1-m_z^2)r^2 dr > 0$.

## Characteristics of equilibrium PBS's

It follows from the foregoing that at the action of a periodic magnetic field only the PBS with a frequency specified by this field may arise spontaneously. This is different from the phase transition discussed in [1], where the frequency and, accordingly, the configuration of arising PBS are not defined.

In Figs.1–3, the configurations of the PBS for several frequencies of precession at $h = 0.998$ are presented. For each frequency, there is the solution of the equation (12) with corresponding PBS that precesses in substratum of uniform precession of the bulk crystal. Beside such solution, there is the solution that corresponds to homogeneous precession of magnetic moments, without soliton. Such a solution at $\omega = 1.2 \times 10^{-3}$ is shown in Fig.1.

Note that the maximum frequency of PBS at $h = 0.998$ without periodic field equals $\omega_{res} = 2 \times 10^{-3}$ and corresponds to magnetic resonance in a metastable state. In Fig.1 and in the following examples, the same parameters as in the [1] article ($M_0 = 0.5 \times 10^{14}\dfrac{eV}{Oe\,cm^3} \cong 80\,Oe$, $K_1 = 1000\,Oe$, $\alpha = \alpha_z = 3\times 10^{-10}\,Oe\,cm^2$) are used and furthermore, $H_\perp = 15 \times 10^{-3}\,Oe$, $\kappa = 5 \times 10^{-4}$.

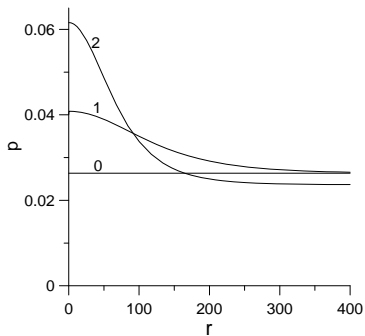

Fig.1 Configurations of solutions of the (12) equation for $h = 0.998$, $h_\perp = 15 \times 10^{-6}$: number 1 is at $\omega = 1.2 \times 10^{-3}$, and number 2 is at $\omega = 1.18 \times 10^{-3}$. Number 0 is the solution for homogeneous precession, without PBS, at $\omega = 1.2 \times 10^{-3}$.

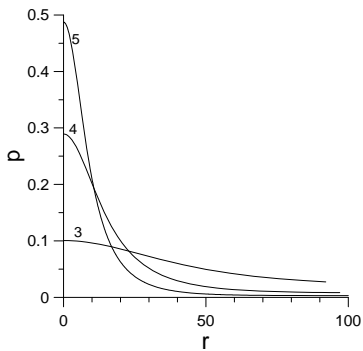

Fig.2 The same as in Fig.1, for the following frequencies: number 3 is if $\omega = 1.11 \times 10^{-3}$, number 4 is for $\omega = 0$, and number 5 is for $\omega = -3 \times 10^{-3}$.

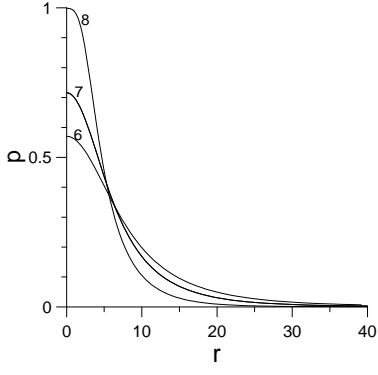

Fig.3 The same as in Fig.1, for the following frequencies: number 6 is for $\omega = -0.005$, number 7 is for $\omega = -0.01$, and number 8 is for $\omega = -0.03$.

In Figs.4 and 5, the frequency dependencies for main parameters of PBS at $h = 0.998$ are presented: the energy $E_s = 4\pi M_0 K_1^{-0.5} \alpha^{1.5} \int e(r) r^2 dr$ (in correspondence with (7)); the amplitude $p_{sm} = (p_{im} - p_0)$ (here $p_0$ is the amplitude of uniform precession); and radius of PBS is $r_{0.5}$.

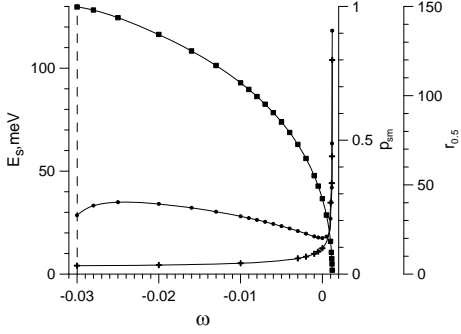

Fig.4 Frequency dependencies of energy (full circles), amplitude $p_{sm}$ (full squares), and radius $r_{0.5}$ (crosses) of PBS for $h = 0.998$, $h_\perp = 1.5 \times 10^{-5}$. Here and further, the utmost values of parameters of equilibrium PBS's are noted by a dotted line.

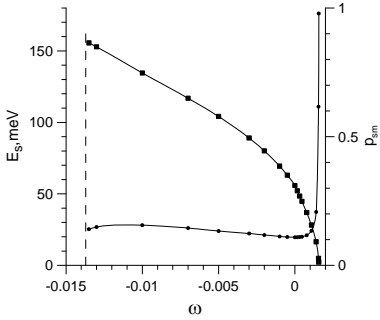

Fig.5 Frequency dependencies of energy and amplitude $p_{sm}$ of PBS for $h = 0.998$, $h_\perp = 0.6 \times 10^{-5}$.

As can be seen from (11), if $p_{sm} \leq 1$, the equilibrium state of PBS is possible only at the (under) following condition:

$$|\omega| p_{sm} \leq h_\perp / \kappa . \tag{14}$$

Frequency range of equilibrium PBS increases with the increase of periodic field amplitude. For $h = 0.998$, if $h_\perp = 1.5 \times 10^{-5}$, $\kappa = 5 \times 10^{-4}$, condition $-\omega p_{sm} = h_\perp / \kappa$ corresponds to $\omega_{min} \cong -0.03$. If $h_\perp = 0.6 \times 10^{-5}$, then $\omega_{min} \cong -0.0137$ (see Figs. 4 and 5).

The possibility of equilibrium PBS is connected with the fact, that during their precession, the magnetic moments, lag by phase behind precession of periodic field. In the Figs. 6 and 7, the frequency dependency of the angle of such a delay in center of PBS, i.e. $\beta_0(\omega) = \beta(r=0)$, is shown. Besides, in Fig.6 the curve of corresponding exponential factor, that defines temperature dependency of the PBS probability, is presented. The probability of PBS origin decreases sharply at the increase of PBS energy.

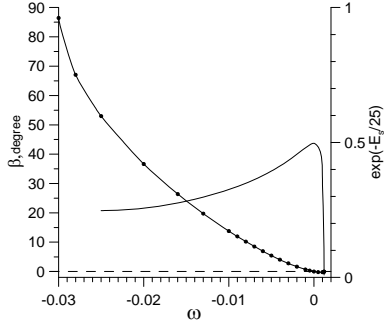

Fig.6 The frequency dependencies of the $\beta_0$ angle if $h = 0.998$, $h_\perp = 1.5 \times 10^{-5}$, and corresponding exponential factor at $T = 300K$.

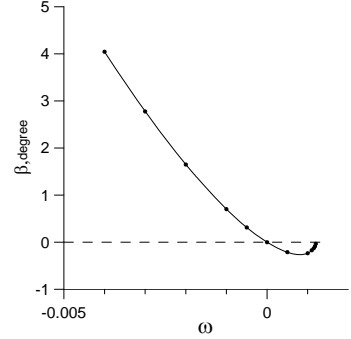

Fig.7 The same frequency dependency of the $\beta_0$ angle as in Fig.6 in the range near $\omega = 0$.

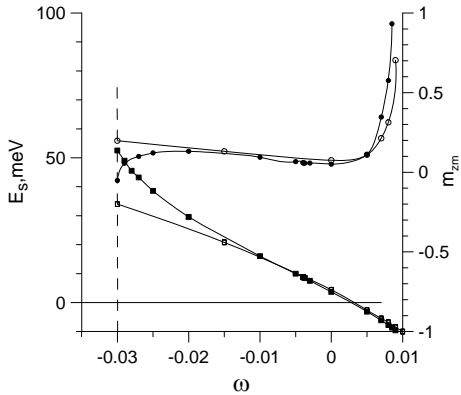

Fig.8 The frequency dependencies of energy and amplitude for equilibrium PBS if $h = 0.99$, $h_\perp = 1.5 \times 10^{-5}$. For comparison, these dependencies for $h_\perp = 0$ case are shown too (by open circles for energy and open squares for the amplitude). The energy $E = 0$ is for the initial homogeneous metastable state.

If the magnetic field $h_\perp$ is enough large, the equilibrium PBS's with amplitude $m_{zm} > 0$ are possible. It is obviously in such cases, the condition for equilibrium PBS's can be written as:

$$\omega \geq -h_\perp/\kappa. \qquad (15)$$

In Figs. 8 and 9, two examples of such case for $h = 0.99$ are presented. As seen in the second example, in Fig.9, the equilibrium PBS's are possible with the negative energy and at $E_s = 0$, i.e. in the bifurcation point b. The configuration of such equilibrium PBS is shown in Fig.10.

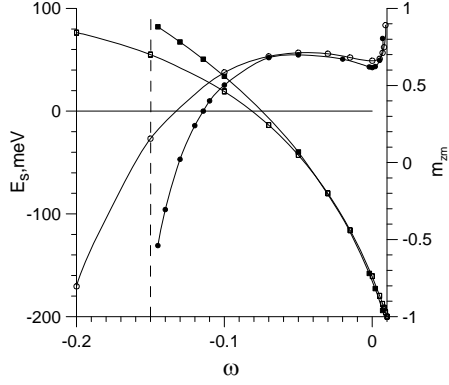

Fig.9 Frequency dependencies of energy and amplitude for equilibrium PBS if $h = 0.99$, $h_\perp = 7.5 \times 10^{-5}$. For comparison, these dependencies for $h_\perp = 0$ case are shown too.

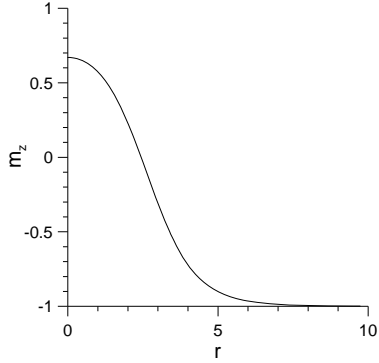

Fig.10 Configuration of equilibrium PBS at $E_s \cong 0$. Here $h = 0.99$, $h_\perp = 7.5 \times 10^{-5}$, $\omega = -0.1143$.

The probability of the equilibrium PBS's near the bifurcation point where $E_s = 0$ increases sharply, as it is shown in Fig.11.

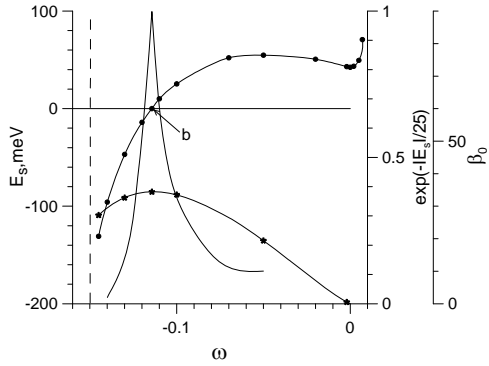

Fig.11 Frequency dependencies of the energy, angle $\beta_0$ (noted by asterisks) and exponential temperature factor for equilibrium PBS's at $h = 0.99$, $h_\perp = 7.5 \times 10^{-5}$. The bifurcation point is noted as **b**.

## Conclusions

Periodic magnetic field acting perpendicular to the axis of easy magnetization, i.e. perpendicular to the main magnetic field, leads to a significant change in the nature of the precessing magnetic solitons and their role in the kinetics of the phase transition:

1. Near the boundary of the existence of the metastable state, only PBS with precession frequency equals the frequency of the external field can occur. The frequency of the external field specifies uniquely of other characteristics of PBS (amplitude, configuration, energy) too.

2. Arising PBS's are equilibrium, i.e. reducing their energy due to dissipative processes is fully compensated by the influx of energy from an external periodic field.

3. We can consider the equilibrium PBS's as "dissipative structures" [2, 3], when the increase of entropy connected with energy dissipation is compensated by a negative flow of entropy due to the action of an external periodic field.

4. At the phase transition in a ferromagnet, the dissipative structures in the form of equilibrium PBS's can be originated not only in the presence of the bifurcation point, but also in a more general case: when the energy of PBS state is small enough (for example, $E_s \cong \kappa_B T$).

5. The compensation of energy and entropy is accompanied by the fact that PBS during its precession delays by phase behind the precession of a periodic field. The magnitude of this delay depends on the radius and the maximum delay is at the center of PBS.

6. On the side of the negative values of frequencies, amplitude and frequency of precession of equilibrium PBS's are defined by the following relations: $-\omega\sqrt{1-m_{zm}^2} = h_\perp/\kappa$ if $m_{zm} < 0$, and $-\omega = h_\perp/\kappa$ if $m_{zm} > 0$.